\documentclass[12pt]{article}
\usepackage{times}
\usepackage{geometry}
\usepackage[utf8]{inputenc}
\usepackage{enumitem,amssymb}
\usepackage{ragged2e}
\usepackage[hyphens,spaces,obeyspaces]{url}
\urldef{\footurl}\url{https://www.insidehighered.com/news/2019/05/23/feds-release-broader-data-socioeconomic-status-and-college-enrollment-and-completion}
\usepackage{caption} 
\usepackage{subcaption} 
\usepackage{hyperref} 

\usepackage{natbib} 
\usepackage{multirow}
\usepackage{graphicx}
\newlist{thematic}{itemize}{8}
\setlist[thematic]{label=$\square$}
\usepackage{pifont}
\usepackage{color}

\usepackage{ulem}

\begin{document}
\raggedright
\LARGE
Astro2020 White Paper \linebreak
Enabling Terminal Master's Degrees as a Step Towards a Ph.D.  \linebreak
\normalsize

  
\textbf{Principal Author:}

Name:	Michael B. Lund
 \linebreak						
Institution:  Caltech/IPAC-NExScI
 \linebreak
Email: Mike.Lund@gmail.com
 \linebreak
 
\textbf{Co-authors:} Savannah R. Jacklin\footnote[1]{Department of Physics and Astronomy, Vanderbilt University, Nashville, TN}, David Ciardi\footnote[2]{Caltech/IPAC-NExScI, Pasadena, CA}
  \linebreak
\textbf{Endorsers:}  
Joseph A. Barranco\footnote[3]{Chair, Department of Physics \& Astronomy, San Francisco State University, San Francisco, CA}, 
Andreas Bill\footnote[4]{Chair, Department of Physics \& Astronomy, California State University Long Beach, Long Beach, CA}, 
Damian J. Christian\footnote[5]{Chair, Department of Physics \& Astronomy, California State University Northridge, Northridge, CA}, 
Robert Fisher\footnote[6]{M.S. Physics Graduate Program Director, Department of Physics, University of Massachusetts Dartmouth, Dartmouth, MA}, 
David Garrison\footnote[7]{Interim Associate Dean for the College of Science and Engineering, Physics Department, University of Houston Clear Lake, Houston, TX} 
Donald J. Jacobs\footnote[8]{Department of Physics and Optical Science, University of North Carolina at Charlotte, Charlotte, NC},
Gaurav Khanna\footnote[9]{Graduate Program Director, Engineering \& Applied Science Ph.D. Program, University of Massachusetts Dartmouth, Dartmouth, MA}, 
Eric C. Landahl\footnote[10]{Graduate Program Director, Department of Physics, DePaul University, Chicago, IL}, 
Andrew Layden\footnote[11]{Chair, Department of Physics \& Astronomy, Bowling Green State University, Bowling Green, OH}, 
Joshua Pepper\footnote[12]{Department of Physics, Lehigh University, Bethlehem, PA},
Seth Redfield\footnote[13]{Chair, Astronomy Department, Wesleyan University, Middletown, CT}, 
Joseph E. Rodriguez\footnote[14]{Center for Astrophysics | Harvard \& Smithsonian, Cambridge, MA},
Eric L. Sandquist\footnote[15]{Chair, Department of Astronomy, San Diego State University, San Diego, CA}, 
Marc S. Seigar\footnote[16]{Associate Dean of Swenson College of Science and Engineering, Department of Physics \& Astronomy, University of Minnesota Duluth, Duluth, MN} 
\linebreak
\textbf{Abstract  (optional):}
Earning any advanced degree in physics or astronomy is an arduous process and major accomplishment. However, not every journey to the Ph.D. is paved equally. Every year, there are hundreds of students who earn terminal master's degrees in physics and astronomy in the United States. A master's degree on its own is sufficient qualification for many good careers, but for a portion of these students, the master's degree is not the final step in graduate education. When students with master's degrees decide to continue their education and are accepted to Ph.D. programs, they often find that their credits do not transfer and that they will be required to re-do large portions of their master's degree at their new Ph.D.-granting institution. Here we discuss the need for gathering more data to understand both the different pathways to a Ph.D. and the students that choose each route. We also discuss some of the challenges faced by students that earn a master's degree before beginning a Ph.D. program. As students in the physical sciences that complete a master's and a Ph.D. at different schools take over 2 years longer to reach a Ph.D. than students that get both degrees from the same school, we suggest steps that can be taken to help these students succeed in a timely manner.

\pagebreak
\section{Introduction}
The road to a Ph.D. can take more than one pathway. In the US, the most common route is completing a bachelor's program and then directly entering a Ph.D. program, generally with the option to get a master's in passing. However, hundreds of students graduate each year with terminal master's degrees in physics and astronomy, and a subset of these students going on to either astronomy Ph.D. programs or physics Ph.D. programs with astronomy research.

Students that complete a master's degree before entering a Ph.D. program at a different university represent several different paths. This includes students that entered stand-alone terminal master's programs at universities that do not extend to a Ph.D., students that are leaving a Ph.D. program with a MS to find a more compatible Ph.D. program, and students that have taken part in bridge programs designed to help them reach Ph.D. programs from non-traditional backgrounds (such as the Fisk-Vanderbilt Master's-to-PhD Bridge Program \citep{Stassun2010} and the APS Physics Bridge program \citep{Beckford2016}). While each of these paths have their own challenges towards the shared goal of a Ph.D. in physics or astronomy, only the traditional path is well-defined, and difficulties faced during non-traditional paths are often not addressed.

\subsection{Master's degrees in STEM}\label{sec:MS_STEM}
For Ph.D.s in science, 29\% of graduates earn a master's degree at an institution other than where they completed their Ph.D.\citep{Lange2006}. Of those that earned a master's degree en route to a STEM Ph.D., 12\% earned their master's degree at master's colleges (schools that offer undergraduate degrees and also award at least 20 master's degrees per year). This number undergoes a statistically significant rise to 18\% for Native American and African American students \citep{Lange2006}. Women completing bachelor's degrees in mathematics and non-biological sciences and then enrolling in graduate education were also less likely to eventually be enrolled in Ph.D. programs \citep{Clune2001}, with 31\% of men reaching a Ph.D. program compared to only 17\% of women. \textbf{A significant portion of graduate students start in a master's program rather than a Ph.D. program, and underrepresented groups are more likely to follow this path. Easing the transition from master's programs to Ph.D. programs may then also help to increase the number of students from traditionally underrepresented groups that advance to Ph.D. programs.\footnote{For a more specific discussion of issues facing underrepresented minorities in astronomy, see Section 1.5 of Astro2020 white paper \citet{Moravec2019}}}

\subsection{Terminal Master's Degrees and Physics and Astronomy}
The currently available data on terminal master's degrees in either physics or astronomy departments specifically is relatively limited. Most existing data is broadly focused on master's degrees as a whole or all STEM fields. The American Institute of Physics (AIP) does collect some useful data on graduate programs. Their yearly surveys of physics and astronomy programs do collect some demographic data on students, but while this does include race and gender, it does not include socioeconomic data such as parent education, or family income, or the need to find additional income outside of grad school. The AIP also conducts a survey of first-year graduate students and one insight this provides is the types of support for graduate students, although only in their first year. Finally, an assessment of students graduating with a Ph.D. can provide some of the educational history of graduating students (such as graduate degrees earned at other schools). This data doesn't include the level of rigorous longitudinal studies on physics and astronomy students that would be needed to look at attrition rates of graduate students, including how socioeconomic status or earning a terminal master's degree influences attrition rates. \textbf{The existing data collected about physics and astronomy graduate students is valuable, but is an incomplete demographic picture. Reliable completion/attrition rates are also not included in this data.}  
\bigbreak
In order to highlight some differences between Ph.D. programs and terminal master's programs specifically in physics and astronomy, we use the undergraduate student population to provide insight into the sorts of backgrounds of students being served by each university. The AIP provides two yearly surveys of university departments, one for astronomy departments \citep{Nicholson2018_ast} and one for physics departments \citep{Nicholson2018_phy}. The data collected includes the highest degree offered by the department, as well as how many students have graduated with each degree offered. As many physics departments include astronomy when a separate astronomy department is not offered, we combine these two department rosters to get a single list of universities with the highest degree offered. Our analysis represents 190 schools where the highest degree in physics and/or astronomy is a Ph.D., and 55 schools where the highest degree is an MS. For the 2016-2017 year, this represents 556 students that exited with a master's degree at universities that grant Ph.D.s, and another 283 students that completed master's degrees at universities where this was the highest degree available (15 programs did not provide data). These schools are then matched against the U.S. Department of Education's College Scorecards\footnote{\url{https://collegescorecard.ed.gov}  \citep{CollegeScorecard2015}}. Here three properties of the 2016 undergraduate student body are examined.
\begin{figure*}[!htb]
  \begin{center}
      \includegraphics[width=0.6\textwidth]{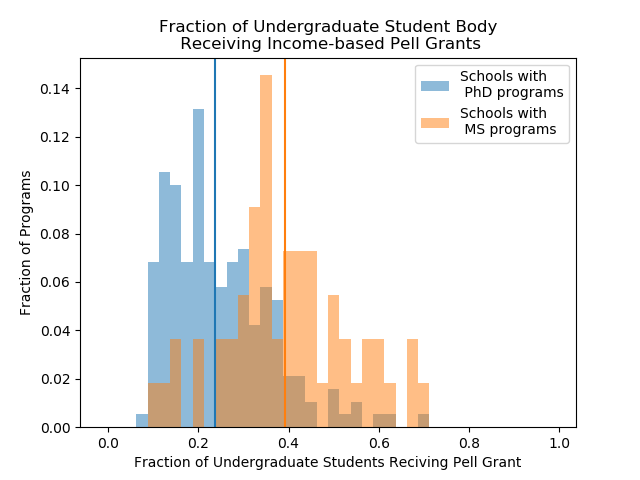}
  \end{center}
  \caption{Fraction of undergraduate student body receiving federal Pell grants designed for low-income students. Vertical lines show the median values for Ph.D. (blue) and MS (orange) granting institutions, or 24\% and 39\% respectively.}
  \label{fig:Frac_Pell}
\end{figure*}
\begin{itemize}
\item In Figure~\ref{fig:Frac_Pell}, we show the fraction of undergraduate students receiving income-based Pell grants at both Ph.D. and master's granting institutions. Pell grants are government grants awarded to U.S. citizen undergraduate students who demonstrate financial need based on family income and cost of tuition. 
 In line with the Department of Education, we use the fraction of students receiving Pell grants as a proxy for the fraction of students coming from lower-income backgrounds.

The average school that only offers a master's degree has about 40\% of its undergraduate students meet the necessary criteria to receive a Pell grant, whereas only $\sim\%20$ students at Ph.D. granting institutions meet the same criteria, suggesting a notable difference in the fraction of the student body that is lower-income.

\item Figure~\ref{fig:Frac_PT} displays the fraction of students categorized as part-time students. Universities that only offer terminal master's degrees have about twice as many undergraduate students that are part-time. A gap also appears fpor physics and astronomy students at the graduate level; according to AIP data, in their first year in grad school, 8\% of students in master's programs were part-time, compared to only 2\% of students in Ph.D. programs. This statistic, in part, represents college students who need to work to support themselves while simultaneously taking classes. \textbf{Schools where the highest degree offered is a master's degree tend to serve student populations that are lower-income, and more likely to be attending part-time.}

\begin{figure*}[!htb]
  \begin{center}
      \includegraphics[width=0.6\textwidth]{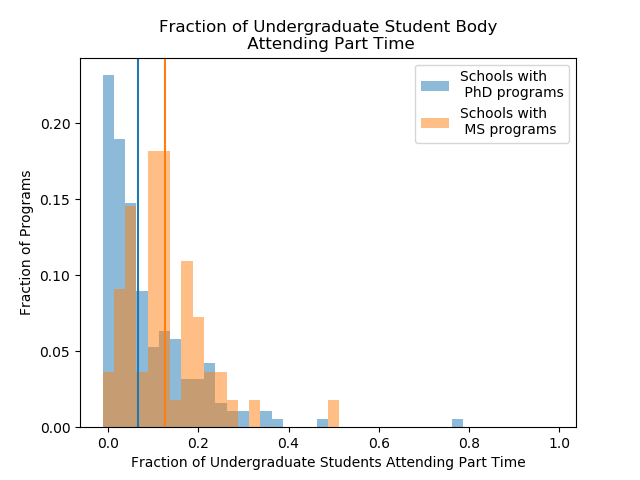}
  \end{center}
  \caption{Fraction of undergraduate student body considered part-time students. Vertical lines show the median values for Ph.D. (blue) and MS (orange) granting institutions, or 7\% and 13\% respectively.}
  \label{fig:Frac_PT}
\end{figure*}
\item Figure~\ref{fig:Frac_FG} shows the fraction of first generation college students. At institutions that offer a Ph.D., about one in four undergraduates is a first generation college student. At schools that offer only a terminal master's degree, this number increases to more than one in three. \textbf{Schools that offer only a terminal master's degree tend to serve undergraduate student populations that are more likely to be the first generation in their family to attend college.}

\begin{figure*}[!htb]
  \begin{center}
      \includegraphics[width=0.6\textwidth]{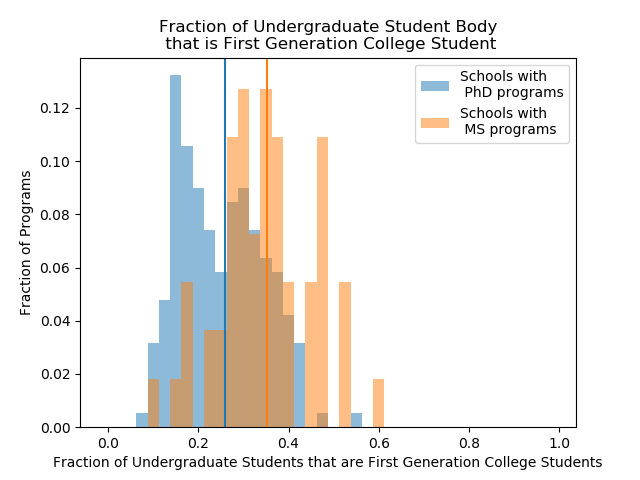}
  \end{center}
  \caption{Fraction of undergraduate student body that are first generation college students. Vertical lines show the median values for Ph.D. (blue) and MS (orange) granting institutions, or 26\% and 35\% respectively.}
  \label{fig:Frac_FG}
\end{figure*}
\end{itemize}
AIP surveys of grad students in their first year can also provide a potential proxy for socioeconomic status of graduate students by looking at the number of first-year graduate students who took community college classes between high school graduation and enrollment in an undergraduate degree. It is not surprising that students from lower socioeconomic backgrounds are more likely to enroll in community colleges after high school for both cost and educational reasons. Approximately 60\% of the high school graduates from the lowest socioeconomic quintile enrolled in community colleges if they went on to post-secondary education, while less than 20\% of the highest socioeconomic quintile students did the same \footnote{\footurl}. From the AIP surveys, in departments that offer only a physics master's degree, 30\% of students had enrolled in community college after high school. In physics departments that offer a Ph.D., this number drops to 17\%, and in departments offering astronomy graduate degrees (where almost all are Ph.D. granting institutions), only 12\% attended some community college. \textbf{Students that have taken coursework at community colleges (which have lower tuition rates than public universities) are a much larger share of programs with a terminal master's degree than of programs with a Ph.D.}

\subsection{Concerns for Master's students}\label{sec:concerns}
Graduate students that earn a master's degree en route to a Ph.D. see longer times to graduation, with the gap varying by field. In a study of University of California Ph.D. programs in the physical sciences, a student that earns a Ph.D. without earning a separate master's degree in the process graduates in an average of 5.5 years. The time to graduation increases to 6.5 years for students that earn a master's degree at the same school as their Ph.D., and to 8.4 years if they earned a master's degree at a different institution than their Ph.D. \citep{Nerad1991}. A similar gap is present in the AIP data for the median full-time equivalent years of graduate study for Ph.D. students that hadn't transferred at 6 years, while for transferring Ph.D. students, it is 5.5 years at their Ph.D. granting institution, and an additional 2 years at their previous institution, for a gap of 1.5 years. \textbf{Students that complete a master's degree and go on to a different Ph.D. program take $\sim2$ years longer in graduate school than students that enter a Ph.D. program directly.}  
\bigbreak
This difference in how long it takes to reach the Ph.D. can represent a significant opportunity cost, as additional years in grad school are years that are not spent earning a higher salary or being able to save money for future expenses. Ph.D. programs generally have financial support of roughly \$15,000 to \$35,000 per year\footnote{From voluntarily submitted data on \url{http://www.phdstipends.com}}.
However, a bachelor's graduate in physics has a median starting salary of \$69,900 directly out of college for the class of 2018 according to surveys of employers\citep{NACE2018}, and a median of \$50,000 according to AIP data on the starting salaries of the classes of 2015 and 2016.
Ph.D. graduates from the classes of 2015 and 2016 earned median starting salaries ranging from \$50,000 in academic postdoctoral positions to \$105,000 for positions in private industry (again from AIP surveys). \textbf{Depending on the salary and gap considered, an extra 1.5 years in grad school can represent roughly \$20,000 to \$135,000 in lost earnings, and an extra 3 years in grad school represents \$40,000 to \$270,000 in lost earnings.}

\bigbreak
Lost earnings are further compounded by the difference in funding for master's students vs. Ph.D. students. \textbf{The physics bachelor's graduating classes of 2009 and 2010 were surveyed in their first year of grad school, and while only 1\% of those in Ph.D. programs were self-funded, 25\% of those in master's programs were self-funded \citep{LangerTesfaye2012}.}

\section{Suggestions for the Decadal Survey}
We recommend two sets of goals for the Decadal Survey to consider with respect to future discussions of master's students that continue on to Ph.D. programs. The first set of goals are tied to the quality of the data available for graduate students. This relies, firstly, on departments providing much of this data about their own programs and students, and secondly on an organization to collect this data. A natural fit would be for increased support to the AIP's Statistical Research Center which is already gathering data on graduate students and graduate programs in physics and astronomy. As graduate programs tend to be small, there should also be some discussion as to how to collect all this data and make it accessible in a way that maintains the privacy of individual students.

\begin{itemize}
\item While the AIP already does collect some demographic data (particularly race and gender) for graduate students, it would be greatly beneficial to have a more complete set of demographic data in both master's and Ph.D. programs, particularly socioeconomic data. This will allow differences between master's and Ph.D. programs more broadly to be examined, such as the extent that lower-income or first generation college students are being represented in the graduate populations. This can also include full-time vs part-time status, and if they are otherwise employed.
\item A greater set of data on graduate students' educational history as they enter and exit graduate programs should also be collected, including longitudinal studies. This would include elements such as any past degrees, years earned, and what subjects they were in. Some of the questions that this data (in conjunction with demographic data) should be able to answer:
  \begin{itemize}
      \item What fraction of physics and astronomy bachelor's graduates go directly into Ph.D. programs and what fraction go directly into terminal master's programs? 
      \item What fraction of students entering a Ph.D. program are leaving another school's Ph.D. program (representing transfers either with or without a master's degree)?
      \item What fraction of graduate students that enter Ph.D. program successfully complete it? How does this vary for students that enter a Ph.D. program directly, complete a terminal master's degree before a Ph.D. program, or that transfer between Ph.D. programs?
  \end{itemize}
\item Students leaving a graduate program should have some data collected, such as what their loan burden is. This will allow an exploration on what the financial burden is on those completing Ph.D. programs and how this is influenced by a terminal master's degree.
\item Data should be collected on funding for graduate students (including extent of support for tuition or health care) in a transparent fashion to help students make informed decisions about their expected out-of-pocket costs at a graduate school level. Reasonable cost-of-living estimates should also be included.
\end{itemize}

The second set of goals are tied to how graduate programs prepare students for terminal master's degrees, as well as how graduate programs assess and support student that have already earned a master's degree in physics, astronomy, or a related field.

\begin{itemize}
\item Ph.D. programs should be encouraged to be more accepting of transferred course credits from master's programs. Reducing the amount of coursework that must be redone can shorten the total time in graduate school, reducing both the total tuition costs and the current opportunity cost that comes with spending additional years in graduate school.
\item To support the above goal, it would be worth investigating how to ensure that master's degree programs provide students with comparable preparation to the first eyars of a Ph.D. program. One idea is the development of a subject-oriented accreditation system that would ensure a consistent level of rigor for graduate programs in physics and astronomy. Such a program could streamline the grad school process by making it easier to transfer credits, which in turn would reduce time-to-Ph.D. for students that earn a terminal master's degree. Such a system would need to be flexible enough to fit departments that vary in both student/faculty size and range of fields covered. It would also likely need to come with more support provided to programs that have limited resources. We do note that a proposal for accreditation for undergraduate physics programs has already been explored by the American Physical Society (APS) and the American Association of Physics Teachers (AAPT), and both organizations independently reached the conclusion that accreditation was not the appropriate course to take.
\linebreak
A different approach would be to instead upon the work currently being done by the APS and AAPT. Instead of accreditation, the APS and AAPT have decided to create a guide to help undergraduate physics programs improve themselves and have access to resources and information on best practices as part of a project called the Effective Practices for Physics Programs (EP3)\footnote{https://www.aps.org/programs/education/ep3/}. The guide is currently in development (expected release in Spring of 2020), and will provide departments with steps for self-assessment and evidence-based strategies for a wide range of goals, such as advising and mentoring, improving curricula, research opportunities, and departmental leadership. A similar approach that is designed for specifically for physics and astronomy graduate programs could then provide departments the guidance to improve without some of the concerns raised by accreditation. A key component of such a program would also need to be providing the support for smaller programs to take advantage of these resources, such as travel to workshops or providing outside assistance to help departments conduct self-assessments and develop plans for improvement. This program could be created by extending the APS/AAPT program to include graduate programs, or through establishing a similar program by another organization, like the AAS or the AIP.
\item As a significant portion of master's students are self-funded and more likely to come from schools serving lower-income populations, reducing the application cost of Ph.D. programs can make applying to Ph.D. programs easier and more affordable. Some suggestions include accepting digital copies of official transcripts or test scores and only requesting official copies upon admission (so that applying to multiple schools is more affordable), and removing mandatory requirements for tests like the GRE that have high associated costs even though their utility has been questioned \citep{Miller2014, Sealy2019}.
\item Much of this section addresses how to make relatively modest changes to the current US system, but a much larger approach would be to move towards thinking of a Ph.D. as a distinct program 3-4 years long that follows a master's degree rather than a longer program that must include a master's degree within it. Currently, one of the options that provides the shortest road to a Ph.D. for those completing a terminal master's degree is to look at schools internationally, where they can immediately begin work towards their Ph.D. and where a master's degree is a requirement for entry into the program.
\end{itemize}

\section*{Acknowledgements}
We would like to thank Patrick Mulvey of the AIP Statistical Research Center for providing significant help in understanding the data currently collected on physics and astronomy students and departments. Any unpublished data attributed to the AIP Statistical Research Center in this paper was provided via personal communication.

Software: matplotlib \citep{Hunter2007}, numpy \citep{Oliphant2006}, pandas \citep{Mckinney2011}


\bibliographystyle{apalike}
\bibliography{main}

\begin{thebibliography}{}

\bibitem[Beckford, 2016]{Beckford2016}
Beckford, B. (2016).
\newblock {Bridge Programs as an approach to improving diversity in physics}.
\newblock {\em PoS}, ICHEP2016:313.

\bibitem[Clune et~al., 2001]{Clune2001}
Clune, M., Nu{\~{n}}ez, A., and Choy, S. (2001).
\newblock {Competing Choices: Men's and Women's Paths}.
\newblock Technical Report March, U.S. Department of Education. National Center
  for Education Statistics, Washington, D.C.

\bibitem[{Executive Office of the President of the United States},
  2015]{CollegeScorecard2015}
{Executive Office of the President of the United States} (2015).
\newblock {Using federal data to measure and improve the performance of U.S.
  institutions of higher education}.

\bibitem[Hunter, 2007]{Hunter2007}
Hunter, J.~D. (2007).
\newblock {Matplotlib: A 2D graphics environment}.
\newblock {\em Computing in Science and Engineering}, 9(3):99--104.

\bibitem[Lange, 2006]{Lange2006}
Lange, S.~E. (2006).
\newblock {\em {The master degree: A critical transition in STEM doctoral
  education}}.
\newblock PhD thesis, University of Washington.

\bibitem[{Langer Tesfaye} and Mulvey, 2012]{LangerTesfaye2012}
{Langer Tesfaye}, C. and Mulvey, P. (2012).
\newblock {Physics Bachelor's One Year Later}.
\newblock {\em AIP Statistical Research Center}, (June).

\bibitem[Mckinney, 2011]{Mckinney2011}
Mckinney, W. (2011).
\newblock {pandas : a Foundational Python Library for Data Analysis and
  Statistics}.
\newblock (January 2011).

\bibitem[Miller and Stassun, 2014]{Miller2014}
Miller, C. and Stassun, K. (2014).
\newblock {A Test that Fails}.
\newblock {\em Nature}, 510:303.

\bibitem[{Moravec} et~al., 2019]{Moravec2019}
{Moravec}, E., {Czekala}, I., and {Follette}, K. (2019).
\newblock {Astro2020 APC White Paper: The Early Career Perspective on the
  Coming Decade, Astrophysics Career Paths, and the Decadal Survey Process}.
\newblock {\em arXiv e-prints}, page arXiv:1907.01676.

\bibitem[{National Association of Colleges and Employers}, 2018]{NACE2018}
{National Association of Colleges and Employers} (2018).
\newblock {Salary Survey: Starting Salary Projections for Class of 2018 New
  College Graduates}.
\newblock Technical report.

\bibitem[Nerad and Cerny, 1991]{Nerad1991}
Nerad, M. and Cerny, J. (1991).
\newblock {From Facts to Action: Expanding the Education Role of the Graduate
  Division}.
\newblock {\em Communicator, Special Edition – May}.

\bibitem[Nicholson and Mulvey, 2018a]{Nicholson2018_ast}
Nicholson, S. and Mulvey, P.~J. (2018a).
\newblock Roster of astronomy departments with enrollment and degree data,
  2017: Results from the 2017 survey of enrollments and degrees.
\newblock Technical report, AIP Statistical Research Center.

\bibitem[Nicholson and Mulvey, 2018b]{Nicholson2018_phy}
Nicholson, S. and Mulvey, P.~J. (2018b).
\newblock Roster of physics departments with enrollment and degree data, 2017:
  Results from the 2017 survey of enrollments and degrees.
\newblock Technical report, AIP Statistical Research Center.

\bibitem[Oliphant, 2006]{Oliphant2006}
Oliphant, T.~E. (2006).
\newblock {\em {Guide to NumPy}}.
\newblock Trelgol Publishing.

\bibitem[Sealy et~al., 2019]{Sealy2019}
Sealy, L., Saunders, C., Blume, J., and Chalkley, R. (2019).
\newblock The gre over the entire range of scores lacks predictive ability for
  phd outcomes in the biomedical sciences.
\newblock {\em PLOS ONE}, 14(3):1--17.

\bibitem[Stassun et~al., 2010]{Stassun2010}
Stassun, K.~G., Burger, A., and Lange, S.~E. (2010).
\newblock {The Fisk-Vanderbilt Masters-to-PhD Bridge Program: A Model for
  Broadening Participation of Underrepresented Groups in the Physical Sciences
  through Effective Partnerships with Minority-Serving Institutions}.
\newblock {\em Journal of Geoscience Education}, 58(3):135--144.

\end{thebibliography}

\end{document}